\documentstyle[12pt,epsf,epsfig]{article}
\def\beq{\begin{equation}}
\def\eeq{\end{equation}}
\def\bea{\begin{eqnarray}}
\def\eea{\end{eqnarray}}
\def\bq{\begin{quote}}
\def\eq{\end{quote}}

\def\NP{{\it Nucl.Phys.} }
\def\PL{{\it Phys.Lett.} }
\def\PR{{\it Phys.Rev.} }

\def\gappeq{\mathrel{\rlap {\raise.5ex\hbox{$>$}}
{\lower.5ex\hbox{$\sim$}}}}

\def\lappeq{\mathrel{\rlap{\raise.5ex\hbox{$<$}}
{\lower.5ex\hbox{$\sim$}}}}

\begin{document}
\pagestyle{empty}
\begin{flushright}
{CERN-TH/95-119}
\end{flushright}
\vspace*{5mm}
\begin{center}
{\bf $k_\bot$ FACTORIZATION vs. RENORMALIZATION GROUP:}\\
{\bf A SMALL-$x$ CONSISTENCY ARGUMENT} $^{*)}$ \\
\vspace*{1cm}
{\bf Marcello Ciafaloni}$^{+)}$ \\
\vspace{0.3cm}
Theoretical Physics Division, CERN \\
CH - 1211 Geneva 23 \\
\vspace*{2cm}
{\bf Abstract} \\ \end{center}
\vspace*{5mm}
\noindent
I investigate, at leading twist level, the consistency of the BFKL
equation
with the renormalization group, when next-to-leading log-$x$ terms are
included in the quark-sea channel of the anomalous dimension matrix. By
use of
$k_\bot$-factorization, I find that, besides next-to-leading small $x$
resummation formulae, a leading, $x$-dependent redefinition of initial
quarks
and gluons is needed. Its interpretation and phenomenological relevance
are
briefly discussed.

\vspace*{5cm}

\noindent  \rule[.1in]{16.5cm}{.002in}

\noindent
$^{*)}$ Work supported in part by E.C. ``Human Capital and Mobility"
contract \# ERBCHRXCT 930357, and by M.U.R.S.T. (Italy).

\noindent
$^{+)}$ On sabbatical leave of absence from Dipartimento di Fisica,
Universit\`a di Firenze, and INFN, Sezione di Firenze, Italy.
\vspace*{0.5cm}

\begin{flushleft}
CERN-TH/95-119 \\
May 1995
\end{flushleft}
\vfill\eject

\setcounter{page}{1}
\pagestyle{plain}

Recent experiments \cite{aaa} on deep inelastic scattering (DIS) at
small
Bjorken $x$ call for a better understanding of the theoretical
predictions on
the proton structure functions, in order to yield, hopefully, some
unambiguous
explanation of their observed small-$x$ rise and of the related scaling
violations.

	Although present approaches to fit the HERA data are mostly based on
QCD
\cite{bb} and most of them successful, different authors follow
different lines
of thought and different procedures, according to whether they emphasize
the
perturbative $Q^2$-evolution based on the GLAP equations \cite{cc}, or
instead
the high-energy, small-$x$ evolution based on the BFKL equation
\cite{dd}.

On the other hand, the systematic use of $k_\bot$-factorization
\cite{ee},\cite{ff}, in order to combine the BFKL equation with the
renormalization group, has shown that the high-energy structure implies
resummation formulas for coefficient functions \cite{ee} and anomalous
dimensions \cite{ggg} in the effective small-$x$ coupling $\alpha_s\log
x$.
This suggests that the two points of view mentioned before may indeed be
equivalent, with the important -- and fruitful -- consequence that
perturbation
theory is recast in a resummed form, so as to explain the HERA data
\cite{hh}.

That is not all, however, because it is usually thought that the BFKL
equation
(and generalizations of it) are able to predict, for large enough
scales, the
small-$x$ behaviour of the cross-section, i.e., the ``perturbative
Pomeron",
even when there is no $Q^2$ evolution at all \cite{jj}, and thus the
anomalous
dimension is not really relevant. Is this a sign that there is some
essential
difference?

The purpose of this note is to push further the consistency point of
view in
experiments with one large scale $Q^2$ (of DIS or Drell-Yan type) and at
leading twist level, i.e., by neglecting subleading powers of
$Q^2_0/Q^2$,
where $Q_0 > \Lambda$ defines the boundary of the perturbative approach
$(\alpha_s(Q^2_0) \lappeq 1)$.

By analysing the BFKL Green's functions with quark loops up to
next-to-leading
(NL) $\log x$ level, I will show that indeed they are equivalent to the
GLAP
ones with properly resummed anomalous dimensions, but they also involve
leading, $x$-dependent redefinition of initial quarks and gluons. It
will
appear that such a redefinition carries the small-$x$ evolution at fixed
scale
mentioned before, even if its precise form is questionable in a
perturbative
approach.

An interesting by-product of the above analysis will be a simple
calculation of
the (subleading) quark anomalous dimensions $\gamma^N_{qg},
\gamma^N_{qq}$ in a
slightly different factorization scheme from the ones of MS-type used by
Catani
and Hautmann \cite{ggg}.

I will consider, for a given moment index $N$, the Green's function
matrix
$G^N_{ba}(Q^2,Q^2_0)$ of the BFKL equation in the flavour singlet sector
defined by
\beq
G^N_{ba}(Q^2,Q^2_0) = \int^{Q^2}_0 d^2k ~~{\cal F}^N_{ba}
({\bf k},Q^2_0)~,\quad (a,b = q,g)~,
\label{1}
\eeq
where the ${\cal F}_{ba}$'s are unintegrated structure functions of
parton $b$
in parton $a$, satisfying small-$x$ equations of the BFKL type, to be
specified
shortly.

In the gluon channel, I shall take ${\cal F}^N_{gg} = {\cal F}_N
({\bf k},Q_0)$, satisfying the usual BFKL equation \cite{kk} in four
dimensions, with fixed initial gluon virtuality $k^2_a = Q^2_0$, i.e.,
\beq
{\cal F}_N({\bf k},Q_0) = \delta ({\bf k}^2-Q^2_0) +
{\bar\alpha_s\over N-1}~~\int~{d^2q\over \pi {\bf q}^2}~~
\left[{\cal F}_N ({\bf k} +{\bf q}, Q_0) -  \Theta (k-q) ~{\cal
F}_N ({\bf k},Q_0)\right]
\label{2}
\eeq
where $\bar\alpha_s = {3\alpha_s\over\pi}$ is the strong coupling
constant.

The solution of Eq. (\ref{2}) admits an integral representation in the
anomalous
dimension plane which, inserted in the definition (\ref{1}), yields
\beq
G^N_{gg} = \int^{{1\over 2}+i\infty}_{{1\over 2}-i\infty}~~{d\gamma\over
2\pi
i}~~ {(Q^2/Q^2_0)^\gamma \over \gamma (1- {\bar\alpha_s\over N-1}~\chi
(\gamma))^{-1}}~,
\label{3}
\eeq
where the eigenvalue function
\beq
\chi (\gamma ) = 2\psi (1) - \psi (\gamma ) - \psi (1-\gamma )
\label{4}
\eeq
implicitly defines the perturbative branch of the gluon anomalous
dimension by
$$
1 = {\bar\alpha_s\over N-1} ~~\chi(\gamma
)~,\phantom{xxxxxxxxxxxxxxxxxxxx}
\eqno{(5a)}
$$
$$
\gamma^N_{gg} = \gamma_N (\alpha_s) = {\bar\alpha_s\over N-1} + 2 \zeta
(3)
\left({\bar\alpha_s\over N-1}\right)^4 + \ldots \quad .
\eqno{(5b)}
$$
\addtocounter{equation}{1}

For $Q^2 > Q^2_0$, $G^N$ is determined by the $\gamma$ poles given by
(5a) in
the left-hand $\gamma$ plane and, at leading twist level, by $\gamma_N$,
in the
form
\beq
G^N_{gg} = K_N(\alpha_s)~~\left({Q^2\over
Q^2_0}\right)^{\gamma_N(\alpha_s)}~~\left( 1 + O\left({Q^2_0\over
Q^2}\right)^p
\right)
\label{6}
\eeq
where the coefficient
\beq
K_N (\alpha_s) = \left[ -\gamma_N {\bar\alpha_s\over N-1} ~~\chi^\prime
(\gamma_N)\right]^{-1} = 1 + 6\zeta (3) \left({\bar\alpha_s\over
N-1}\right)^3 +
\ldots \label{7}
\eeq
will be further discussed in the following.

Note that, at this stage, I keep $\alpha_s$ frozen, because the running
of
$\alpha_s$ mixes with other NL log effects, which are still under
investigation
\cite{lll}. Of course, once a factorization of the R.G. type is
established, the
running of $\alpha_s$ will be restored.

I shall consider, instead, NL effects due to quark loops, because they
couple
directly to the photon. The final quark channel is in fact defined in
terms of
the structure function $F_2$ \footnote{This was called the DIS scheme in
Ref.
\cite{ggg}. However, the present definition of gluons is different [cf.
Eq.
(18)].}
\beq
F^N_{2a} = \sum_f~~e^2_f~~G^N_{qa}
\label{8}
\eeq
and, by $k_\bot$ factorization, its evolution is given \cite{ggg} in
terms of
the off-shell $\gamma g \rightarrow \bar qq$ cross-section
$\hat\sigma_2$ (Fig. 1)
\beq
{\partial\over \partial\log Q^2}~~G_{qa}^N~~(Q^2,Q^2_0) =
\int~~d^2k~~{\partial\over
\partial\log Q^2}~~\hat\sigma_2~\left({k^2\over Q^2}\right) ~~{\cal
F}_{ga}
(k^2,Q^2_0)~,
\label{9}
\eeq
where the lowest-order expression of $\hat\sigma_2$ was given in Ref.
\cite{ee}.

Note that the definition (1) and Eq. (9)  carry $k_\bot$- integrations
up to
$k^2 = 0$. Nevertheless, for fixed $\alpha_s$, they are well defined,
because,
according to Eq. (\ref{3}), the region $k^2 < Q^2_0$ is governed by the
r.h.
plane anomalous dimensions $(1-\gamma_N)$ and higher, and is hence
automatically suppressed. It appears, therefore, that fixing the initial
parton
virtuality at $Q_0$ automatically regulates all intermediate collinear
singularities, unlike what happens in (on-shell) dimensional subtraction
schemes. Furthermore, the overall collinear singularity will be
factorized in
$Q_0$-dependent powers, as done in Eq. (6) in the gluon channel. I will
call this factorization procedure the $Q_0$-scheme, to distinguish it
from the
ones of MS-type, used in Ref. \cite{ggg}.

The most important consequence of introducing quark loops comes from Eq.
(\ref{9}), where the explicit expression of $\hat\sigma_2$ \cite{ee} was
used
in Ref. \cite{ggg} to provide the NL resummed formula for the
off-diagonal
anomalous dimension $\gamma_{qg}$. In fact, by setting $a = g$ and by
performing the $k_\bot$-integration with the help of Eq. (\ref{3}), one
gets
$$
{\partial\over\partial\log Q^2}~~G^N_{qg}~(Q^2,Q^2_0) \equiv \dot
G^N_{qg} = 2N_f h_2
(\gamma_N (\alpha_s))~~(G^N_{gg} + O(Q^2_0 / Q^2)^p)
\eqno{(10a)}
$$
where
$$
h_2(\gamma ) = {\alpha_s\over 3\pi}~~{1 + {3\over 2} \gamma
(1-\gamma)\over 1 -
{2\over 3} \gamma}~~{[\Gamma (1-\gamma) \Gamma (1+\gamma)]^3\over \Gamma
(2-2\gamma)
\Gamma (2+2\gamma]}~,
\eqno{(10b)}
$$
\addtocounter{equation}{1}
and $N_f$ is the number of quark flavours. On the other hand, the
customary
GLAP evolution equations would give, to NL accuracy \cite{mm}:
$$
\dot G^N_{qg} = \gamma^N_{qa}~G^N_{ag} = \gamma^N_{qg}~G^N_{gg} +
O(\alpha^2_s)~,
\eqno{(11a)}
$$
$$
\dot G^N_{qq} = \gamma^N_{qa}~G^N_{aq} =
\gamma^N_{qg}~G^N_{gq} + \gamma^N_{qq} + O(\alpha^2_s)~.
\eqno{(11b)}
$$
\addtocounter{equation}{1}
Therefore, by comparing (11a) with (10a), we obtain
\beq
\gamma^N_{qg}(\alpha_s) = 2N_f~h_2 (\gamma_N(\alpha_s))~.
\label{12}
\eeq

This determination of $\gamma_{qg}$ in the $Q_0$ scheme differs from
Ref.
\cite{ggg} by the absence of the so-called $R_N$ factor [analogous to
$K_N$ in Eq.
(\ref{6})], which arises \cite{nn} from regularizing and factorizing the
intermediate collinear singularities in MS-type schemes. I shall further
comment on this point later on.

By then introducing initial quarks (see Fig. 1), and by neglecting
internal
quark loops at NL level, I obtain integral representations for all
$G_{ab}$'s,
as follows
$$
\dot G^N_{ab} = \int^{{1\over 2}+i\infty}_{{1\over 2}-i\infty} ~~
{d\gamma\over 2\pi
i}~~{(Q^2/Q^2_0)^\gamma\over 1 - {\bar\alpha_s\over N-1} \chi
(\gamma)}~~ h_a(\gamma
)~ k_b(\gamma )~,
\eqno{(13a)}
$$
where
$$
h_g = k_g = 1,~~h_q = {2N_f\over \gamma}~h_2(\gamma)~,~~~k_q = {C_F\over
C_A}~{1\over \gamma}~{\bar\alpha_s\over N-1}~,
\eqno{(13b)}
$$
with $C_F = 4/3$ and $C_A = 3$.
\addtocounter{equation}{1}

It is now not difficult to evaluate Eq. (13) for $Q^2 \gg Q^2_0$, at
leading
twist level and NL log $x$ accuracy. Since the flavour singlet anomalous
dimension matrix has two eigenvalues, one close to $\gamma_N$, and the
other
next-to-leading, I will keep in Eq. (13) all terms coming from poles at
$\gamma
= \gamma_N$ and $\gamma = 0$ which, after some algebra, yield the
following
results.
\begin{enumerate}
\item[(a)] The $G_{ab}$'s {\it do~satisfy} the GLAP-type equations
(11), i.e., in matrix form,
\beq
\dot G_N (Q^2,Q^2_0) = \Gamma_N (\alpha_s) G_N(Q^2,Q^2_0)
\label{14}
\eeq
with $\gamma_{gg} \; (\gamma_{qg})$ given in Eq. (5) [Eq. (12)]
respectively, and
\beq
\gamma^N_{gq} = {C_F\over C_A}~\gamma_N~,~~~\gamma^N_{qq} = {C_F\over
C_A}~\left(\gamma^N_{qg} - \gamma^{(1)}_{qg}\right)~,
\label{15}
\eeq
where $\gamma^{(1)}_{qg} = 2N_f\alpha_s/3\pi$ is the lowest-order
expression.

Equations (5), (12) and (15) yield the promised resummation formulas in
this
scheme.
\item[(b)] The normalization of the $G_{ab}$'s {\it is~not} the
same as the corresponding GLAP Green's function matrix, which for frozen
$\alpha_s$ is $F_N = (Q^2/Q^2_0)^{\Gamma_N}$, but differs from it by a
constant, finite, $N$-dependent matrix $K^N$
\beq
G_N(Q^2,Q^2_0) = (Q^2/Q^2_0)^{\Gamma_N}~K^N(\alpha_s)~,
\label{16}
\eeq
where
\bea
K^N_{gg} &=& K_N~,~~~K^N_{gq} = {C_F\over
C_A}~~\left({\gamma^{(1)}_N\over\gamma_N}~K_N -1\right)~,\nonumber \\
K^N_{qg} &=& {\gamma^N_{qg}\over \gamma_N}~K_N -
{\gamma^{(1)}_{qg}\over\gamma^{(1)}_N}~,~~~
K^N_{qq} = 1 + {C_F\over C_A}~\left(
{\gamma^{(1)}_N\gamma^N_{qg}\over\gamma^2_N}~K_N
- {13\over 6}~\gamma^{(1)}_{qg}\right)~,
\label{17}
\eea
and $\gamma^{(1)}_N \equiv \bar\alpha_s/(N-1)$.
\end{enumerate}

Several comments are in order. First, one should not be too surprised to
find a
different determination of NL anomalous dimensions in the $Q_0$ scheme.
It is
easy to check that $\Gamma_N^{(Q_0)}$ differs from $\Gamma_N^{(DIS)}$ in
Ref.
\cite{ggg} by a redefinition of the gluon density, as follows:
\beq
\Gamma_N^{(Q_0)} = a_N^{\Gamma_L}~\Gamma_N^{(DIS)}~a_N^{-\Gamma_L}~,
\label{18}
\eeq
where $\Gamma_L$ is the common leading part (provided by $\gamma_{gg}$
and
$\gamma_{gq})$ and $a_N$ is a (properly chosen) scale-changing
parameter.

It is, however, of some interest that the present determination does not
include
the $R_N$ (or $K_N)$ factors, which contain a strong Pomeron singularity
(for
instance, $K_N$ diverges like $(\gamma_N-1/2)^{-1}$ for $N \rightarrow
N_{\bf
P}$, or $\gamma_N\rightarrow 1/2)$. For this reason, the resummed
expression
for $\Gamma^{(Q_0)}$ [Eq. (12)] is expected to be smoother than the one
for
$\Gamma^{(DIS)}$, and may be, as such, more palatable. A firm conclusion
will
be possible when NL contributions to $\gamma^N_{gg}$ [which are changed
by Eq.
(18)] will be available.

More importantly, Eq. (16) clarifies that, if the BFKL equation is to
provide
the same parton densities $f_a(Q^2)$ as GLAP's at the scale $Q^2$, then
the
initial densities must be {\it different}, as follows
\beq
f_a^{\rm GLAP} (Q^2_0) = K^N_{ab}(\alpha_s)~f_b^{\rm BFKL}(Q^2_0)~.
\label{19}
\eeq
The normalization matrix $K^N$ carries the information related to the
small-$x$
evolution at the scale $Q_0$, and in fact contains the strong Pomeron
singularity of the $K_N$ factors mentioned before.

Having checked the evolution equations (11), the running coupling can be
restored in the factorized formula (16), in the normal way
\beq
G^N = \exp \left(
 \int^{\log{Q^2\over\Lambda^2}}
 _{\log
{Q^2_0\over\Lambda^2}}~
dt~\Gamma_N(\alpha_s(t))\right)~~K^N
(\alpha_s(Q^2_0))~,
\label{20}
\eeq
with the usual uncertainty connected with NL terms in the gluon channel.
In
this interpretation, the matrix $K^N$ depends on the strong coupling at
the
lower scale, and thus its precise form (17) will be affected by
higher-twist
unitarity effects \cite{oo} if $\Lambda^2/Q^2_0$ is not too small, and
anyway much
before they become relevant at the scale $Q^2$. Whatever the actual form
of $K^N$,
Eq. (20) teaches us that approaches based on the BFKL (resp. GLAP)
equations,
can only be compared when such redefinition of quarks and gluons is
accounted
for.

To conclude, the information coming from the BFKL equation for single
hard
scale processes can be incorporated in the renormalization group in a
consistent picture. The effects of small-$x$ evolution due to
fluctuations in
$k_\bot$ around $Q_0$ are factorized out and can be incorporated in a
change of
initial conditions. Whether such an ``initial Pomeron" looks like the
BFKL one
or like the one of soft hadronic physics \cite{pp} is a
($Q_0$-dependent) question
open to investigation \cite{oo}, \cite{qq}, which has not been touched
here.

\vspace{1cm}
\noindent
{\bf Acknowledgements}

It is a pleasure to thank Stefano Catani and Francesco Hautmann for
interesting
discussions and suggestions, and Guido Altarelli, Keith Ellis, Stefano
Forte,
Giuseppe Marchesini, Paolo Nason and Graham Ross for helpful
discussions.

\vfill
\eject
\vspace*{6cm}

\begin{figure}
\epsfxsize=15cm\epsfbox{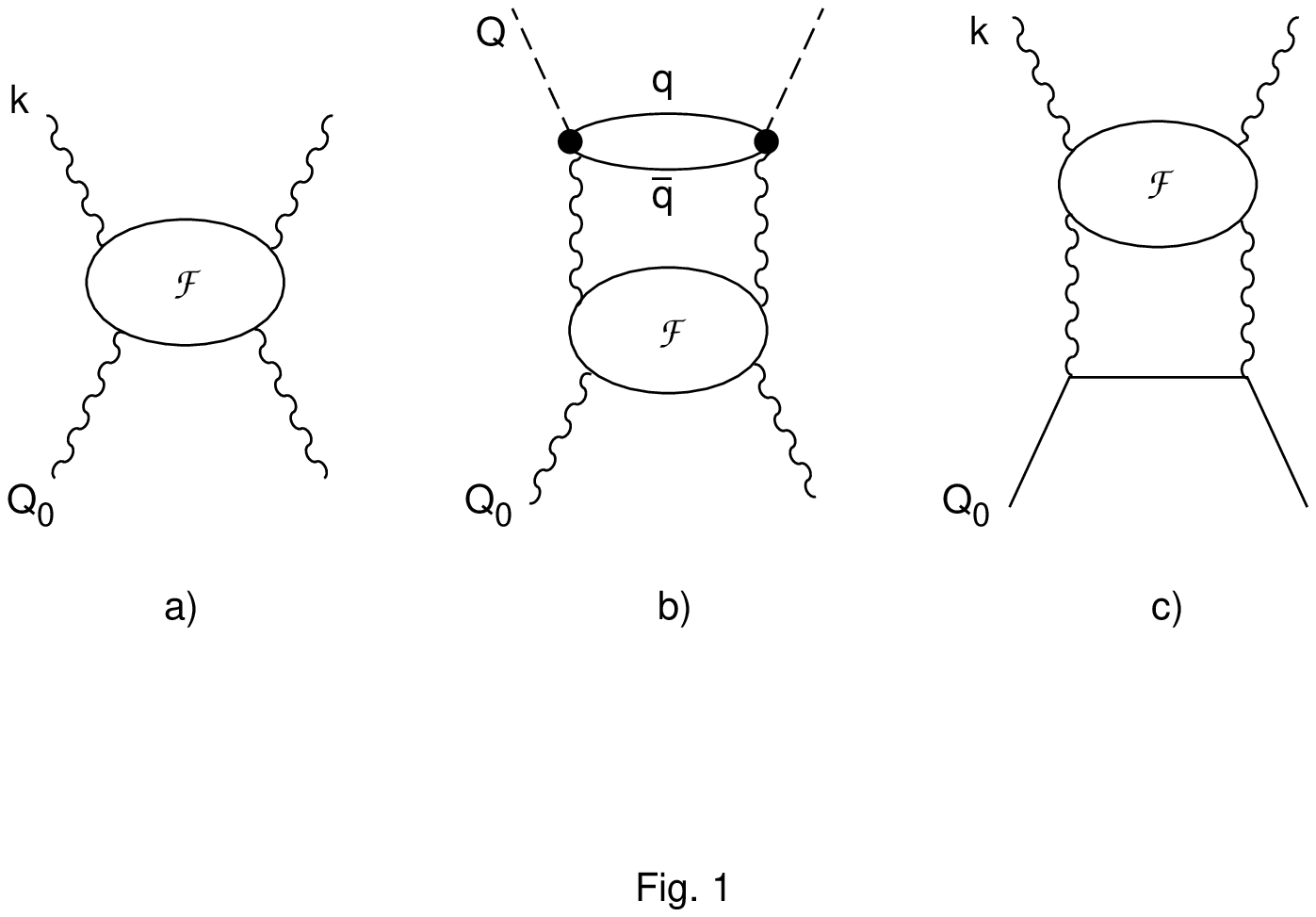}
\end{figure}
\vspace*{1cm}
Diagrammatic definition of (a) the gluon-gluon (b) the quark-gluon and
(c) the
gluon-quark Green's functions up to next-to-leading $\log \; x$ level.
Wavy
(dashed) lines denote (Regge) gluon (photon) exchanges.

\end{document}